\documentclass[prstab]{revtex4}
\usepackage[latin1]{inputenc}
\usepackage{epsfig,amssymb,amsmath}
\usepackage{textcomp}

\usepackage{graphicx}
\newcommand{\comments}[1]{}
\newcommand{\exb}[0]{$E$$\times$$B$ }
\begin{document}

\title{Influence of temperature fluctuations on plasma turbulence investigations with Langmuir probes}

\author{B. Nold$^1$}
\author{T.~T. Ribeiro$^2$}
\author{M. Ramisch$^1$}
\author{Z. Huang$^1$}
\author{H.~W. M\"uller$^2$}
\author{B.~D. Scott$^2$}
\author{U. Stroth$^{1,2}$}
\author{{the ASDEX Upgrade Team}$^2$}

\affiliation{1 Institut für Plasmaforschung, Universität Stuttgart,
70569 Stuttgart, Germany\\
2 Max-Planck-Institut f\"ur Plasmaphysik, EURATOM Association, Garching, Germany}

\date{\today}

\begin{abstract}
The reliability of Langmuir probe measurements for plasma-turbulence
investigations is studied on GEMR gyro-fluid simulations and compared
with results from conditionally sampled $I$-$V$ characteristics as well
as self-emitting probe measurements in the near scrape-off layer of the tokamak ASDEX 
Upgrade. In this region, simulation and experiment consistently show 
coherent in-phase fluctuations in density, plasma potential and also
in electron temperature.
Ion-saturation current measurements turn out to reproduce
density fluctuations quite well. Fluctuations in the floating 
potential, however, are strongly influenced by temperature fluctuations and, hence, 
are strongly distorted compared to the actual plasma potential.
These results suggest that interpreting floating as plasma-potential 
fluctuations while disregarding temperature effects is not justified 
near the separatrix of hot fusion plasmas. Here, floating potential measurements
lead to corrupted results on the \exb dynamics of turbulent
structures in the context of, e.g., turbulent particle and momentum 
transport or instability identification on the basis of 
density-potential phase relations.

\comments{While ion-saturation current measurements turn out to reproduce
density fluctuations quite well, fluctuations in the floating 
potential are governed by those in the temperature and, hence, 
are strongly distorted compared to the actual plasma potential.
These results suggest that interpreting floating as plasma-potential 
fluctuations while disregarding temperature effects is not justified 
in the SOL near the LCFS of hot fusion plasmas. Here, floating potential measurements
lead to corrupted results on the \exb dynamics of turbulent
structures in the context of, e.g., turbulent particle and momentum 
transport or instability identification on the basis of 
density-potential phase relations.
}
\end{abstract}

\maketitle

\section{Introduction}
\label{sec1} 
In the boundary region of magnetically confined fusion plasmas, 
micro instabilities determine a major part of transport losses 
across the separatrix into the scrape-off layer (SOL) and, therefore, 
play a key role for the global confinement quality.
In order to understand the mechanisms degrading or even 
improving confinement via vortex-flow interactions, the
involved dynamical processes are addressed by measurements of,
e.g., correlations between density ($\tilde{n}$) and plasma-potential
($\tilde{\Phi}_{\rm pl}$) fluctuations as well as average radial turbulent 
transport of particles ($\Gamma=\left<\tilde{n}\tilde{v}_r\right>$ 
with radial \exb velocity fluctuations $\tilde{v}_r=\tilde{E}_\theta/B$)
and momentum (perpendicular Reynolds stress $R_s = 
\left<\tilde{v}_\theta\tilde{v}_r\right>$), see e.g.~\cite{Ritz_1987_nf_27_1125,
 Carreras_1996_pop_3_2664, Boedo_2001_pop_8_4826, 
Oost_2003_ppcf_45_621, Xu_2005_ppcf_47_1841, 
Grulke_2006_pop_13_1, XuGS_2009_nf_49_9, Cheng_2010_ppcf_52_055003, Kirk_2011_ppcf_53_3, Mueller_2011_nf_51_073023} among others.

The acquisition of local plasma-potential fluctuations as required for these
studies is mostly based on Langmuir probe measurements.
The measured floating potential of the probe is related to the plasma potential
via 
\begin{equation}
  \label{eq:floating}
  \tilde{\Phi}_{\rm fl} =  \tilde{\Phi}_{\rm pl} - \Lambda   \frac{\tilde{T}_e}{e} 
\end{equation}
with $\Lambda \approx 3$. Electron-temperature fluctuations ($\tilde{T}_e$) are 
rarely available in experiments and usually neglected such that, for simplicity, 
$\tilde{\Phi}_{\rm fl} = \tilde{\Phi}_{\rm pl}$ is used.
The simplifying assumption that temperature fluctuations are negligible has been
proven valid in a toroidally confined low-temperature 
plasma~\cite{Mahdizadeh_2005_47_569}, where temperature gradients were much smaller than the density gradient.
However, in the boundary of several high-temperature plasmas, significant temperature fluctuations have been found by fast sweeping Langmuir 
probes~\cite{Hidalgo_1992_prl_69_1205, Giannone_1994_1_3614, Riccardi_2001_rsi_72_461, Schubert_2007_rsi_78_053505}, tripple probes 
\cite{Riccardi_2001_rsi_72_461, Tsui_1992_rsi_63_4608, Kumar_2003_nf_43_622},
swept double probes~\cite{Xu_1996_pop_3_1022}, 
a harmonic probe technique~\cite{Rudakov_01_rsi_72_1, Boedo_2001_pop_8_4826} 
and Thomson scattering~\cite{Boedo_2001_pop_8_4826, Kurzan_2007_ppcf_49_825}.

In this work, the influence of temperature fluctuations on Langmuir probe 
measurements of density and potential fluctuations in a high-temperature
plasma is investigated. To this end, simulations of plasma-edge turbulence
are carried out with the gyro-fluid code GEMR. 
Synthetic probe data from the simulations are analyzed with respect to 
temperature effects. The results are compared to probe measurements 
near the last closed flux surface (LCFS) in the tokamak ASDEX Upgrade~\cite{Herrmann_2003_fst_44_569}.
For the comparison, two independent sophisticated approaches
are employed to resolve $\tilde{T}_e$ experimentally: one is based on 
self-emitting probe measurements, the other on conditionally sampled probe 
characteristics. It will turn out that 
temperature fluctuations near the LCFS of high-temperature
plasmas strongly distort floating-potential measurements, and an interpretation of
$\tilde{\Phi}_{\rm fl}$ as $\tilde{\Phi}_{\rm pl}$ is tenuous.

This paper is organized as follows:
A short description of the GEMR code is given in Sec.~\ref{sec3}. Section~\ref{sec4} presents the analyzes of the synthetic probe data from the simulations and temperature effects will be identified. The results are compared with 
experimental data from both, self-emitting probe measurements and conditionally sampled probe characteristics, in Secs.~\ref{sec5} and~\ref{sec6}, respectively.
In Sec.~\ref{sec7}, the results are discussed with respect to possible consequences for the interpretation of probe data and a conclusion is given.

\section{GEMR simulations}
\label{sec3}
The three-dimensional gyrofluid
model GEMR~\cite{Zweben_2009_pop_16_082505} solves the first six moments of the gyrokinetic equation (density, parallel velocity, parallel and perpendicular temperature and associated parallel/parallel and perpendicular/parallel heat flux) for ions and electrons. The corresponding conservation
equations are derived using a consistent treatment of the energy
conservation~\cite{Scott_2005_pop_12_102307}. The plasma species are connected through the
quasi-neutrality condition obtained from the gyrokinetic polarization
equation~\cite{Lee_1983_pf_26_556} and the induction obtained from
Amp\`ere's law.
The number of moments retained in the model enables GEMR 
to capture typical tokamak edge regimes, where realistic density and
temperature gradients yield a mixture of drift-Alfv\'en and ion-temperature-gradient (ITG) 
turbulence. Collision frequencies may reach values below the Braginskii collisional regime. 
The gyrofluid formulation guaranties a proper presentation of the unit order finite Larmor radius effects in the drift wave component.

GEMR is a global and not a flux-tube code. It allows the geometry to vary both radially and
poloidally. A circular cross-section with toroidal axisymmetry is assumed. The coordinate system is aligned
with the equilibrium magnetic field to take computational advantage
of the strong spatial anisotropy of magnetised plasmas. 
Global consistency in the angles is ensured, even for toroidally truncated domains \cite{Scott_1998_pop_5_2334}.
The present simulations were performed in a quarter of the torus with a resolution of $128\times512\times16$ in the $x$, $y$ and $s$ directions, respectively. The coordinate system consists of the flux label $x$, the field line label $y$ and the coordinate parallel to the magnetic field $s$. It is linked to the polar coordinates ($r,\theta,\phi$) by the unit Jacobian transformation rules
\begin{equation}
  \label{eq:coords}
  x=2\pi^2R_0r^2
  \qquad
  y_k=\left(q_s\theta-\phi-\alpha_k\right)/\left(2\pi\right)
  \qquad
  s=\theta / \left(2\pi\right)
\end{equation}
with a toroidal shift $\alpha_k(r)$ yield by the shifted metric procedure due to the shearing of the magnetic field \cite{Scott_2001_pop_8_447}.
ASDEX Upgrade parameters were chosen for the tokamak major and minor radii of $R_0=1.65$\,m and $r_0=0.5$\,m, respectively. The safety factor was set to $q_s=4.6$ and the magnetic field to $B=2.6$\,T. The inner half of the radial box corresponds to a $3$\,cm
wide edge region and the outer half corresponds to the SOL, with a
limiter cut located at the bottom of the flux surfaces. 
Within the local approximation of homogeneous normalization parameters, the code gives a consistent description of the turbulence from the confinement
region across the LCFS into the SOL. In
the SOL, the field lines intersect material plates, allowing the existence of
modes with parallel and perpendicular wave numbers $k_\parallel=0$ and $k_\perp\neq 0$, respectively. This changes
substantially the behavior of heat and particle fluxes down the
gradients \cite{Ribeiro_2005_ppcf_47_1657}. 
The SOL is implemented with a change in the parallel boundary conditions according to a linearised Debye sheath model \cite{Ribeiro_2005_ppcf_47_1657, Ribeiro_2008_ppcf_50_055007}, 
consistent with the formulation of GEMR, which retains only quadratic nonlinearities.

The zonal profiles of the
state variables (densities, temperatures and field potentials) are
evolved together with the fluctuations. This accounts for the vigor of SOL turbulence. The linear drive terms are included in the gradients of the evolving profiles. Profile
maintenance is achieved with source/sink zones at
each radial boundary. The zonal components of the densities and
temperatures are feedback dissipated towards an initially specified
profile \cite{Ribeiro_2008_ppcf_50_055007}.
The constant density $n_0$ and temperature $T_0$ used for normalization in the simulations reflect typical ASDEX Upgrade L-mode parameters, namely $T_0=T_i=T_e=60$\,eV and $n_0=n_i=n_e=1.2\times 10^{19}$\,m$^{-3}$. The radial decay lengths of temperature ($L_T$) and density ($L_n$) have been chosen as $L_T = L_n/2 = 3$\,cm. The deuterium ion to electron mass ratio was $m_i/m_e=3670$ and the effective ion charge was $Z_{\rm eff}=2$. The following section compares $4$\,ms of turbulent fluctuations from a saturated phase of the GEMR simulation with simultaneous Langmuir multi-probe measurements at the LCFS in the outboard midplane.

\section{Synthetic-probe results}
\label{sec4}
\begin{figure}
 \centering
 \centerline{\includegraphics[width=0.8\textwidth]{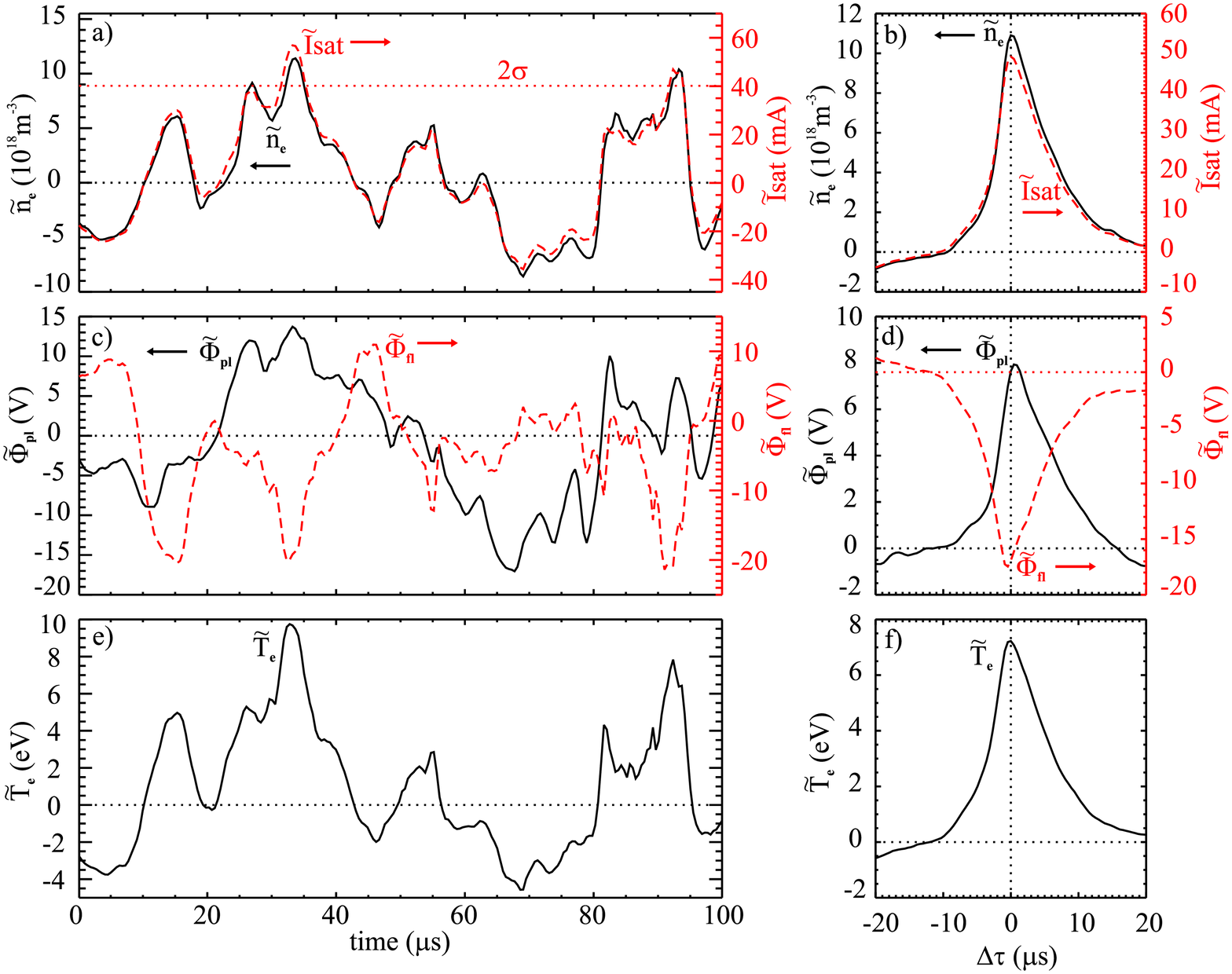}}
  \caption{Comparison of simulated plasma fluctuations with synthetic probe data. Shown are details of the raw time series (left) and the conditionally averaged fluctuations (right) at the LCFS. Top: density and ion-saturation current (dashed) fluctuations; Middle: plasma and floating (dashed) potential fluctuations; Bottom: electron temperature fluctuations from GEMR.} 
  \label{fig3}
\end{figure}

Synthetic Langmuir probes have been implemented in GEMR to extract 
measurements from the simulated data.
The ion-saturation current fluctuations are calculated according to the linearized equation
\begin{equation}
  \label{eq:lin_isat}
  \tilde{I}_{\rm sat}\approx A_p e n_0 c_{s0} 
  		\left(\frac{\tilde{n}_e}{n_0} - \frac{1}{2}\frac{\tilde{T}_{e}}{T_0}\right)
\end{equation}
assuming $T_e = T_i$, to be consistent with the GEMR model. Electron density ($\tilde{n}_e$) and temperature fluctuations ($\tilde{T}_e$) evolve with time, while $n_0$ and $T_0$ are constant normalization parameters. The current is averaged over the parallel projection area ($A_p$) of an ASDEX Upgrade pin probe. The constant ion-sound velocity $c_{s0}$ is calculated using $T_0$. The results do not change significantly if the common expression
  \begin{equation}
  \label{eq:isat}
  I_{\rm sat} = 0.61 A_p e n_e \sqrt{\frac{T_e+T_i}{m_i}}
\end{equation} 
is used instead.  
The floating potential fluctuations are approximated by Eq.~(\ref{eq:floating}) with  $\Lambda = 3.4$ for a cold Langmuir probe in a deuterium plasma. Different models predict similar values for $\Lambda$ \cite{Bissell_1989_pfb_1_5}. The differences are small, especially when compared with the experimental uncertainties concerning secondary electron emission, ion temperature and effective particle collection areas \cite{Schrittwieser_2002_ppcf_44_567}. However, the exact value is not decisive for the nature of turbulent fluctuations, even if the absolute values change.

The left-hand side of Fig.~\ref{fig3} compares the simulated plasma fluctuations $\tilde{n}_e$ (top), $\tilde{\Phi}_{\rm pl}$ (middle) and $\tilde{T}_e$ (bottom) in a $100\,\mu$s time window with the synthetic probe data of $\tilde{I}_{\rm sat}$ and $\tilde{\Phi}_{\rm fl}$. All signals are taken from the LCFS at the outboard midplane of the circular simulation domain. 
Good agreement is found between fluctuations of plasma density and ion-saturation current. This indicates a minor influence of the temperature fluctuations and justifies the common experimental assumption $\tilde{I}_{\rm sat} \propto \tilde{n}_e$ in the simulation.
In contrast, no agreement is found between fluctuations of plasma and floating potential (Fig.~\ref{fig3}c). 
$\tilde{\Phi}_{\rm fl}$ rather follows the inverse temperature 
behavior as shown in Fig.~\ref{fig3}e, which according to 
Eq.~(\ref{eq:floating}) indicates the dominance of $\tilde{T}_e$
in the measured floating potential.

In order to study the correlation between the fluctuating quantities
systematically, the coherent part of the fluctuations is extracted 
by means of the conditional averaging technique~\cite{Johnsen_1987_pf_30_7}.
The ion-saturation current was used as reference signal with a trigger threshold of $2\sigma$ (upper horizontal dashed line in Fig.~\ref{fig3}a), where $\sigma$ is the standard deviation.
The right-hand side of Fig.~\ref{fig3} shows the conditional average of $660$ independent series of fluctuations in density and ion-saturation current (top), plasma and floating potential (middle) and temperature (bottom). 
The average temporal behavior of coherent structures in the synthetic
ion-saturation current and in the density is again found to agree very well.
However, the synthetic floating potential is clearly anti-correlated with the simulated plasma potential fluctuations (Fig.~\ref{fig3}d). This can be attributed to significant coherent temperature structures as depicted in Fig.~\ref{fig3}f, which show up in phase with structures in $\tilde{n}$ and
$\tilde{\Phi}_{\rm pl}$.

The analyses carried out on data from turbulence simulations suggest that
close to the LCFS in high-temperature plasmas, floating-potential fluctuations
measured with probes are strongly affected by temperature fluctuations and,
therefore, do not even qualitatively reflect the plasma potential.
This reflects the experimental situation if two assumptions hold:
(I) The turbulent fluctuations calculated by the gyro-fluid simulation are comparable to plasma fluctuations in the boundary of a real tokamak.
(II) In hot magnetized plasmas, the behavior of Langmuir probes is sufficiently well described by the Debye sheath model.  
Hence, the presence of significant temperature fluctuations would have 
severe consequences for the interpretation of fluctuations from probe 
measurements as will be discussed in Sec.~\ref{sec7}.
In the next two sections, it will be shown that the above assumptions are 
indeed supported by experimental observations.

\section{Emissive-probe measurements}
\label{sec5}
\begin{figure}
 \centering  \centerline{\includegraphics[width=0.8\textwidth]{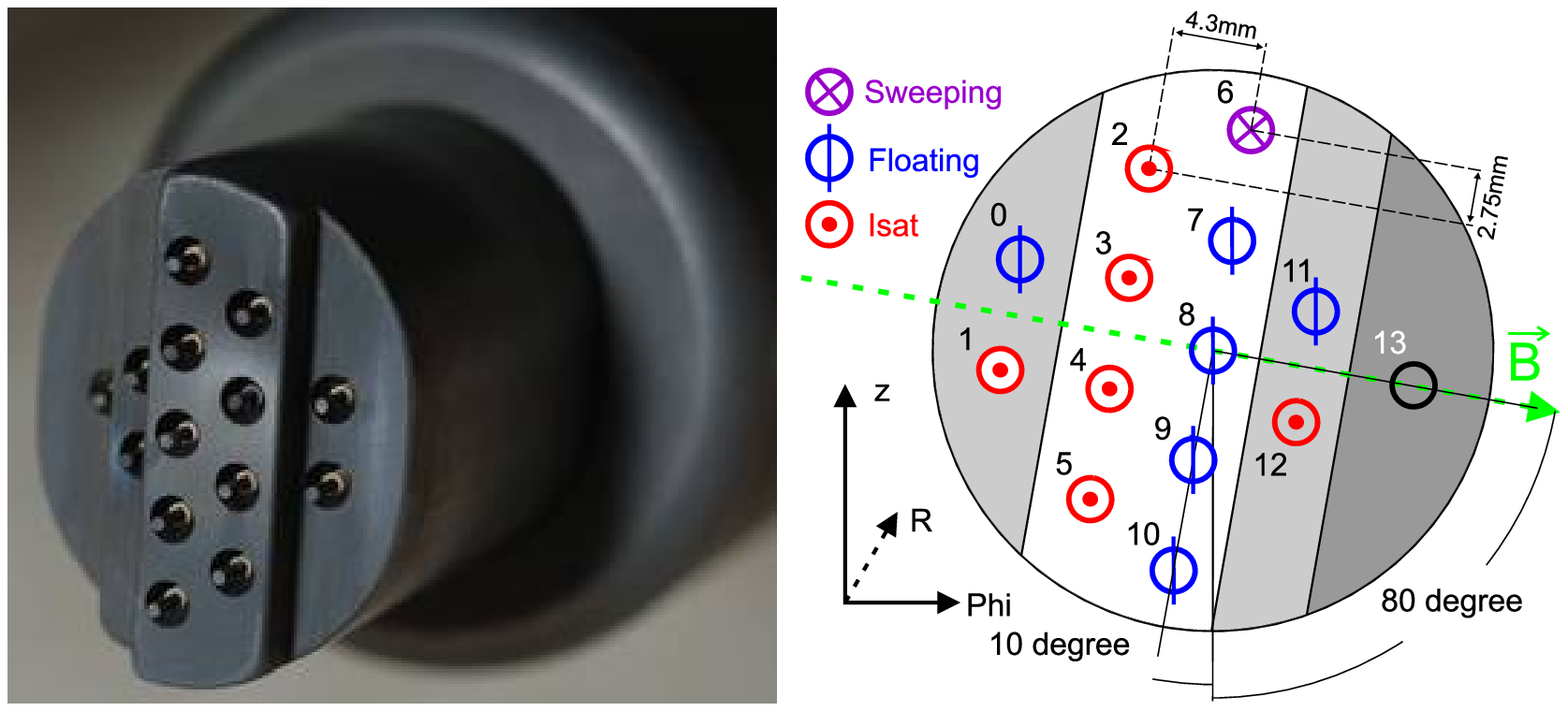}}
	\caption{Photo (left) and sketch (right) of the multi-pin probe head as seen from the plasma in the tokamak ASDEX Upgrade. The shaded areas in the sketch are retracted by $4$\,mm and $8$\,mm (probe 13).}
	\label{fig1}	
\end{figure}
In order to test the validity of the simulation results from the previous section, probe measurements were carried out in 
the tokamak ASDEX Upgrade~\cite{Herrmann_2003_fst_44_569}.
The experiment was conducted in an
L-mode deuterium plasma, kept in lower single null divertor configuration with a constant toroidal magnetic field of $2.5$\,T (clockwise) and a plasma current of $800$\,kA (counter clockwise from top view). The auxiliary electron cyclotron heating power was $600$\,kW and the edge plasma density  $4.0 \times 10^{19}$\,m$^{-3}$.
A Langmuir probe array penetrated the plasma horizontally $0.31$\,m above the outer torus midplane \cite{Nold_2010_ppcf_52_065005}. $8$ free standing carbon pins on the top level measured alternatively floating potential and ion-saturation current, as shown in Fig.~\ref{fig1} (white background). The $I_{sat}$ probes were biased with $-180$\,V and the floating-potential signals were amplified directly at the manipulator exit to minimize low-pass filtering by cable capacity and plasma impedance. The data were sampled at $2$\,MHz with $14$\,bit resolution. For the investigation of only the fast dynamics upon the stationary background,
fluctuations below $10$\,kHz have been digitally removed from the signals.

For comparison with simulation results from the previous section, fluctuations
in the floating and plasma potential, density and temperature are deduced from
the probe signals in two different, independent ways: The first method --
presented in this section --
involves direct measurements of the plasma potential with an emissive probe.
The next section is dedicated to the second method, which is based on 
conditionally sampled probe characteristics.
Both methods are applied to data from the same discharge.
\begin{figure}
 \centering \centerline{\includegraphics[width=0.8\textwidth]{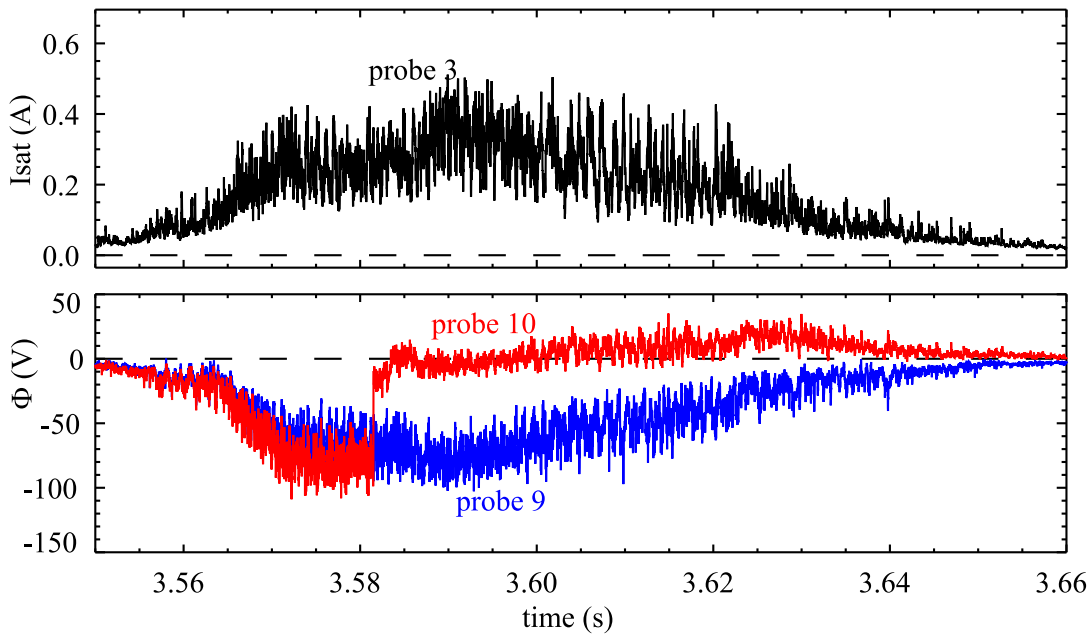}}
  \caption{Measurements of ion-saturation current (top) and potential (bottom) during a stroke of Langmuir probes in ASDEX Upgrade. Due to the high heat load in the plasma, one probe emits electrons during the outward motion and measures a more positive potential (red line).} 
  \label{fig5}
\end{figure}
Figure~\ref{fig5} shows raw data from three Langmuir probes during the second stroke of discharge $\#26530$.
The ion-saturation current ($I_{\rm sat}$, top) increases and the floating potentials ($\Phi_{\rm fl}$, bottom) decrease while the probes approach the LCFS ($t<3.57$\,s). At $t\approx3.582$\,s,  probe number $10$ (red) reaches a critical temperature due to the heat flux from the plasma onto the probe and starts to emit electrons. The emitted electron current is space charge limited after a fast transition within $2\,\mu$s. Now, the self-emitting probe floats at the potential
\begin{equation}
  \label{eq:sepotential}
  \Phi_{\rm se} = \Phi_{\rm pl} - 0.6 \frac{T_e}{e},
\end{equation}
which corresponds to the sheath edge separating sheath from presheath. This was shown in previous investigations with self-emitting probes in ASDEX Upgrade \cite{Rohde_1997_jnm_241_712}. Probe number $9$ is less heated and measures the floating potential of a cold probe
\begin{equation}
  \label{eq:flpotential}
  \Phi_{\rm fl} = \Phi_{\rm pl} - 3.4 \frac{T_e}{e}
\end{equation}
during the entire discharge.

\begin{figure}   
 \centering \centerline{\includegraphics[width=0.8\textwidth]{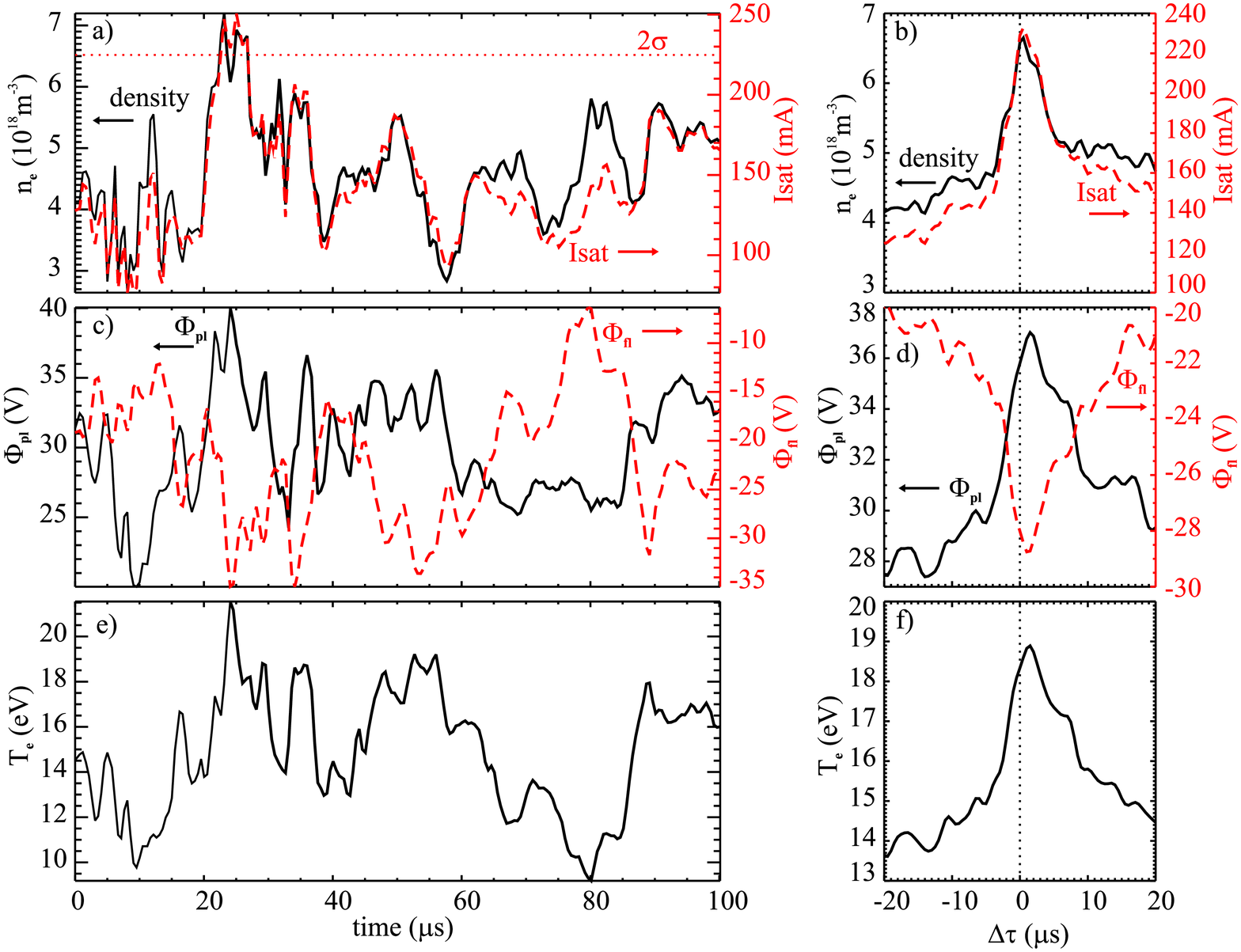}}
  \caption{Comparison of measured plasma fluctuations from cold and hot Langmuir probes in the SOL of ASDEX Upgrade. Detail of time series (left) and conditionally averaged fluctuations (right). Top: density and ion-saturation current (dashed); Middle: plasma and floating (dashed) potentials; Bottom: electron temperature from Eqs.~(\ref{eq:sepotential}) and~(\ref{eq:flpotential}).} 
  \label{fig7}
\end{figure} 
Combining Eqs.~(\ref{eq:sepotential}) and~(\ref{eq:flpotential}) the 
electron-temperature dynamics can be deduced from simultaneous 
measurements of $\Phi_{\rm fl}$ and $\Phi_{\rm se}$ with a temporal resolution of $0.5\,\mu$s. Furthermore, using the result for the temperature and
assuming $T_e \approx T_i$, the density can be calculated according to 
Eq.~(\ref{eq:isat}) from nearby measurements of the ion-saturation current. Similar to Fig.~\ref{fig3}, Fig.~\ref{fig7} shows short fluctuation time traces of the results on the left-hand side and the 
conditional average of $28$ independent large amplitude events with $I_{\rm sat}>2\sigma$ on the right-hand side.
The stationary fluctuations are taken from a $4.1$\,ms time interval at 
$t=3.625\,$s, which corresponds to the radial range $14-18$\,mm outside the 
LCFS.
An anti-correlation is observed between floating and plasma potential (Fig.~\ref{fig7}d), similar to the simulation result in Fig.~\ref{fig3}d. Electron temperature fluctuations with a significant amplitude (Figs.~\ref{fig7}e, f) are found in phase with plasma-potential fluctuations.
Fluctuations in the density are almost coincident with those in $I_{\rm sat}$
(Figs.~\ref{fig7}a, b), i.e.~the temperature fluctuations do not carry 
significant weight. 
In comparison, a good agreement of simulation and experimental results 
is found, which confirms the crucial role of temperature fluctuations
in interpreting probe data.
It can be shown, that the experimental findings are not distorted 
by the probe separation, which introduces additional
phase delays between the signals due to the perpendicular propagation
of turbulent structures. Structures propagating with typically $1$\,km/s 
would show a phase shift of $5\,\mu$s over a distance of $5$\,mm.
This might explain a small phase delay, but does barely account for the 
anti-correlation of the floating and the plasma potential.
Nevertheless, a different method, which does not suffer from 
the probe distances, is used in the next section to verify the
experimental results given here.

\section{Conditional sampling of probe characteristics}
\label{sec6}
\begin{figure}
 \centering \centerline{\includegraphics[width=0.8\textwidth]{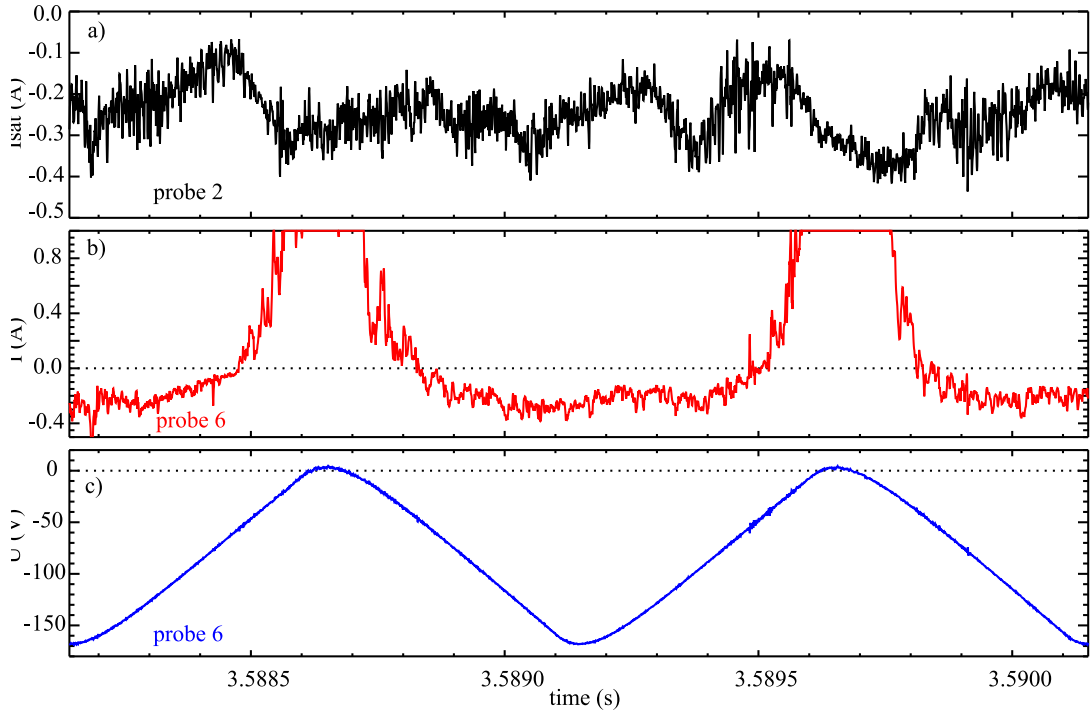}}
  \caption{Example of raw time series for conditional sampling in AUG: Trigger source ion-saturation current (top), current (middle) and applied voltage (bottom) of the swept Langmuir probe.} 
  \label{fig9}
\end{figure}
 In this section, the previous results from emissive probes are verified via 
an alternative method, which resolves fluctuations in potential, electron temperature and
density quasi-instantaneously at the same position. This method has been proposed recently and is based on the conditional sampling of characteristics from a slowly 
swept Langmuir probe~\cite{Furno_2008_pop_15_055903}. Here, it is applied for the first time to
the high-temperature plasma in ASDEX Upgrade.
Figure~\ref{fig9} shows time windows of an ion-saturation current measurement ($I_{\rm sat}$, top) together with the current ($I$, middle) and voltage ($V$, bottom) of a slowly swept Langmuir probe nearby the $I_{\rm sat}$ measurement.
Complete probe characteristics are swept with a rate of $1$\,kHz.
After conditional sampling, a full characteristic is available every microsecond within a short time window around the trigger condition of $1.5\sigma$ amplitude in the $I_{\rm sat}$ signal of probe number $2$.
Around each trigger event at $t_i$, a short $I$-$V$ trace is collected from the neighboring probe number $6$. 
From several trigger events, a complete $I$-$V$ characteristic can be reconstructed at each time lag $\Delta \tau$ with respect to the trigger times $t_i$, since the trigger 
is not phase locked to the sweeping bias voltage. 
In the present case, $510$ trigger events have been detected within a $40$\,ms time interval. During this time, the probe was located in the SOL $8-15$\,mm outside the LCFS. 
The temporal resolution is reduced from $0.5$ to $1\,\mu$s, i.e.~$1020$ $I$-$V$ pairs are combined to one conditionally sampled $I$-$V$ characteristic. This is sufficient to obtain reliable 
fits to the characteristics providing $T_e$, $\Phi_{\rm fl}$ and $I_{\rm sat}$ every microsecond.

\begin{figure}
 \centering \centerline{\includegraphics[width=1.0\textwidth]{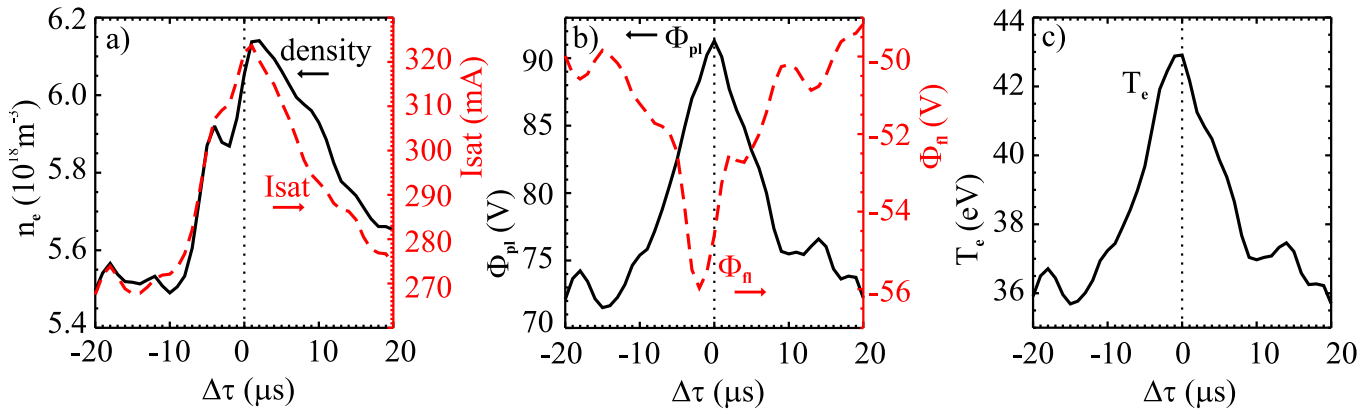}}
  \caption{Results from fitting conditionally sampled $I$-$V$ characteristics close to the LCFS of AUG. Left: density and ion-saturation current (dashed); Middle: plasma and floating potential (dashed); Right: electron temperature from fits to probe characteristics.} 
  \label{fig11}
\end{figure}
Figure~\ref{fig11} shows the fluctuations obtained from fits to the conditionally sampled $I$-$V$ characteristics, together with plasma density and potential calculated according to the Eqs.~(\ref{eq:isat}) ($T_e \approx T_i$) and~(\ref{eq:flpotential}), respectively. 
The results are qualitatively the same as those from emissive-probe measurements:
Density, plasma-potential and temperature fluctuations are essentially in phase.
The ion-saturation current does not substantially deviate from the
density. The floating potential, however, is seriously affected 
by the temperature fluctuations resulting in an anti-correlation
between floating and plasma potential. Moreover, the experimental 
results confirm once more those from the numerical simulations in 
section~\ref{sec4}. This means that (I) the plasma fluctuations in the 
boundary region of ASDEX Upgrade are sufficiently well represented 
by the gyro-fluid simulation and (II) the Debye sheath model is 
appropriate to describe Langmuir probes under the present conditions.
It turns out clearly that great care has to be
taken in interpreting data from Langmuir probes when strong
temperature fluctuations are present.

\section{Discussion and conclusion}
\label{sec7}
The objective of this paper was to study the influence of electron temperature fluctuations on turbulence investigations with Langmuir probes in high-temperature plasmas. 
To this end, gyro-fluid simulations of plasma turbulence were carried 
out for fusion edge plasmas comparable to that in ASDEX Upgrade.
Synthetic Langmuir probes were implemented to mimic the
experimental diagnostics. From the simulations, 
coherent structures in the fluctuations of the electron temperature
could be expected in the SOL close to the LCFS of a real tokamak.
For comparison, Langmuir-probe measurements were carried out in the
edge region of ASDEX Upgrade. The simulation
results could be confirmed in two independent experimental
approaches, one based on self-emitting probe measurements and the
other on conditionally sampled probe characteristics.

In investigations on plasma dynamics, probe measurements of the 
ion-saturation current and the floating-potential fluctuations are widely used
interchangeably for the density and the plasma-potential fluctuations, respectively.
In this work, it was shown that coherent temperature fluctuations
seriously alter the potential measurements.
In simulation and experiment it was found that close to 
the LCFS, coherent density, plasma-potential and 
electron-temperature
fluctuations are essentially in phase. Density fluctuations
are reasonably well represented by the ion-saturation current, i.e. the
influence of $\tilde{T}_e$ is marginal.
Plasma and floating potential, however, show a completely 
different behavior: they are found to be anti-correlated as a 
consequence of the significant $\tilde{T}_e$ contribution. An influence
of $\tilde{T}_e$ on $\tilde{\Phi}_{\rm fl}$ is also observed in a numerical investigation \cite{Gennrich_2011_submitted}.

This has far-reaching consequences for probe based turbulence 
investigations. The phase relation between density and plasma potential
is an indicator for the turbulence-driving instability. It determines
whether a density perturbation is stable or unstable depending on the \exb
advection associated with the corresponding potential perturbation. 
Hence, the cross phase determines the level and even the direction of turbulent particle transport. 
This work shows, that the phase between plasma and
floating potential is governed by the potential-temperature cross phase.
In turn, the floating potential does not reflect the plasma potential.
Deduced quantities like density-potential cross phases and turbulent
transport as well as Reynolds stress involving electric field estimates 
would be distorted if based on floating potential measurements.

For example, the floating potential in Fig.~\ref{fig11}b suggests
that the density is accompanied by a negative monopole 
structure in the potential. In fact, the plasma potential shows a 
positive monopole structure, as expected from the high parallel electron mobility.
As a result the \exb drifts are inverted and indicate an opposite vorticity. From floating-potential measurements, 
the dynamics of coherent potential structures would hence be misinterpreted. 
In particular, the average radial turbulent 
transport $\Gamma=\left<\tilde{n}\tilde{v}_r\right>$ with $\tilde{v}_r=\tilde{E}_\theta/B$ and 
$\tilde{E}_\theta \approx -(\tilde{\Phi}_{\rm fl, 2}-\tilde{\Phi}_{\rm fl,1})/\Delta$
($\Delta$ is the poloidal probe distance) would point into the wrong 
direction. Similarly, estimates of the perpendicular
Reynolds stress ($R_s = \left<\tilde{v}_\theta\tilde{v}_r\right>$) as a measure
of vortex tilting can be affected qualitatively if temperature fluctuations 
deform the topology of the floating compared to the plasma potential.
From the formal construction of $R_s$ which is quadratic in the electric field, it can be seen, however, that 
the polarization alone is not relevant. I.e., $R_s$ measurements near 
the LCFS might be at least qualitatively valid.
Moreover, it is worth noting that in the present study, the fluctuations 
in the SOL near the LCFS of hot fusion plasmas reveal 
drift-wave like features, i.e., a density-potential cross phase close to zero.
In this region, density blobs come along with a monopole potential rather than 
a delayed dipole structure, which would be associated with an interchange 
mechanism behind the blob dynamics. The results are consistent with previous plasma-potential measurements close to the confined plasma of TJ-K~\cite{Happel_2009_prl_102_25},
DIII-D~\cite{Boedo_2001_pop_8_4826} and 
ASDEX Upgrade~\cite{Horacek_2010_nf_50_105001}.
The coincidence of density and electron-temperature fluctuations
reflects the transport of heat into the SOL and is consistent with findings from TEXT, Phaedrus-T, TJ-I, W7-AS and Repute-I~\cite{Hidalgo_1996_cpp_36_139}.
 
In conclusion, plasma fluctuations in the 
SOL near the LCFS of hot plasmas in ASDEX Upgrade are well described
by GEMR gyro-fluid simulations. In this region, measurements and simulation showed 
the existence of coincident, coherent structures in density, potential as well 
as electron-temperature. The temperature fluctuations are large and coherent
enough to significantly distort
floating-potential measurements with Langmuir probes for interpretation as
 plasma-potential fluctuations. At large temperature amplitudes, the distortion
of the floating potential generally depends on the phase and coherence
of the temperature-potential coupling. Hence, great care must be taken
when interpreting floating-potential measurements with respect to, e.g., 
cross-phase relations and turbulent transport of particles or momentum.
Results obtained in the SOL near the LCFS of hot fusion plasmas similar to
ASDEX Upgrade are likely to be corrupted if the effect of temperature fluctuations 
is disregarded. In this work, methods based on self-emitting probe 
measurements or conditionally sampled probe characteristics proved useful to 
test for such temperature effects.

\bibliographystyle{MySty}
\bibliography{fixedlibrary}

\end{document}